\documentclass[11pt]{article}
\usepackage{fullpage}
\usepackage{graphicx}
\usepackage{afterpage}
\usepackage{apalike}
\usepackage{setspace}

\newcommand{\REF}{~\cite}

\begin{document}

\title{{\sc Barnacle:} An Assembly Algorithm for Clone-based Sequences
of Whole Genomes} 
\author{Vicky Choi$^{1,2}$, Martin Farach-Colton$^{1}$\\
$^{1}$Department of Computer Science, Rutgers University, Piscataway, New Jersey
  08854
}
\date{September 5, 2002}
\maketitle

\footnotetext{$^2$Corresponding author. Email: vchoi@cs.duke.edu, Fax: (919)660-6519.}

\doublespacing
\def\abstractname{}

\begin{abstract}
We propose an assembly algorithm {\sc Barnacle} for sequences
generated by the clone-based approach.  We illustrate our approach by
assembling the human genome.  Our novel method abandons the original
physical-mapping-first framework.
As we show, {\sc Barnacle} more
effectively resolves conflicts due to repeated sequences.  The latter
is the main difficulty of the sequence assembly problem. 
Inaddition, we are able to detect inconsistencies in the underlying data.  
We present and compare our results on the December 2001 freeze of the
public working draft of the human genome with  NCBI's assembly (Build 28).

The assembly of December 2001 freeze of the public working draft generated
by {\sc Barnacle} and the source code of {\sc Barnacle} are available at 
(http://www.cs.rutgers.edu/\verb+~+vchoi).

\end{abstract}

Much of the attention and debate within the Human Genome Project has
focussed on two sequencing strategies: {\em whole-genome shotgun
sequencing}, adopted by Celera Genomics\REF{celera-paper}; vs. {\em
hierarchical shotgun sequencing}, also referred to as {\em
clone-based} or {\em BAC-by-BAC}, used by the International Human
Genome Sequencing Consortium (IHGSC)\REF{IHGSC-paper}. 
The data set
generated by the whole-genome shotgun approach consists of a set of
pairs of sequence reads (mates). Each read has average length of
around $500$bp governed by current sequencing technology.  The data
set generated by the clone-based approach consists of a set of
Bacterial Artificial Chromosome (BAC) clones, typically 100-200Kb.
We will use `BAC', `clone' and `BAC clone' interchangeably.
Each BAC clone is individually sequenced and assembled.  In general, a
clone consists of a set of more or less non-overlapping 
{\em fragments}.
A clone is called a
{\em finished clone} if it consists of only one fragment which has
accuracy at least $99.99\%$, otherwise it is a {\em draft
clone}.

The computational problem of reconstructing the genome sequence by
assembling the sequences of the data set is called the {\em sequence
assembly problem}.  The challenge of the problem lies in the
repeat-rich nature of the genome.  For example, the human genome is
more than half repeated sequences, which include large 50-500Kb
duplicated segments with high sequence identity
($\ge\!98\%$)\REF{IHGSC-paper}.
Computationally, the
sequence assembly problems for the two sequencing strategies 
are completely different because the formats of the data sets are
different. 

In this paper, we focus on the assembly problem of sequences generated
by the clone-based approach, with application to the human genome.  In
addition to the repeat problem mentioned above, the assembly
problem of the clone-based approach is further complicated by
laboratory errors, such as chimeras and contaminations; by the draft
quality of the sequences, which are sometimes inaccurate; and by
polymorphism.

The set of BAC clones, finished and draft, produced by IHGSC is called
the {\em public working draft of the human genome}.  It is worthwhile
to point out that the clone-based strategy used is not the originally
proposed ``clone by clone'' strategy, in which a physical map giving
the clone ordering is first produced, and then a minimum tiling set of
clones in the map is selected and individually subjected to shotgun
sequencing.
In practice, the physical map of the clones based on fingerprint
overlaps is constructed concurrently with the
sequencing\REF{IHGSC-paper}.
The ordering of the sequenced clones is either unknown or not
accurately known.

There are two other algorithms for assembling 
the public working draft
of the human genome.  The other assemblers, such as {\sc
ARACHNE}\REF{wgs-paper}, are incomparable because the formats of the input
data are different. The assembly problem of the clone-based approach
 can be viewed as a
second-stage assembler, in which the input fragments 
are the
preassemblies of the shotgun reads of each BAC clone produced by some
assemblers, such as {\sc PHRAP} (Green, unpublished).
One is {\sc GigAssembler}\REF{GP-paper} from UC
Santa Cruz and the other is 
an unpublished 
NCBI's
algorithm 
(http://www.ncbi.nlm.nih.gov/genome/guide/human/).
The fingerprint-based physical map, which was constructed
by manually editting the initial automated fingerprint assembly\REF{IHGMC-paper},
was employed by {\sc GigAssembler} to assist in the sequence assembly,
although the clone ordering was only roughly determined in the
map\REF{GP-paper}. NCBI does not make use of the fingerprint
map
(http://www.ncbi.nlm.nih.gov/genome/guide/HsFAQ.html\#diffassemblies).(Remark:
NCBI's algorithm was said to be modified to use the hand-curated Tiling
Path Files since Build 29.)
Instead, several STS marker maps were used for chromosome
assignments.  Both algorithms are based on greedy approaches to
assemble sequences in which the best overlap is assembled first, where
{\em best} is defined by some score functions used to assign weights
for overlaps. Since the genome is highly repetitive, these score
functions are unlikely to yield the true assembly\REF{care2-paper}.
Without the fingerprint clone contig to have BACs restricted to, the
NCBI algorithm first formulates a Maximal Interval Subgraph (MIS)
problem to obtain a BAC ordering and then assembles the sequences of
overlapping BACs.  This approach is natural given the history of the
Genome Project, with its original physical-mapping-first approach.
However, this top-down approach suffers from the fact that if a BAC is
misplaced, the mistake in the assembly is unrecoverable.

In this paper, we propose an assembler {\sc Barnacle}
for the clone-based
sequences augmented with chromosome-assignment, thus we use the same
input data as NCBI.   Part of the novelty of our approach is to
assemble the genome bottom-up, without making use of fingerprint-based
physical map of the clones.  That is, we use extensive sequence-level
overlaps to suggest BAC overlaps.  
As we will explain in the discussion section, the clone-overlaps inferred from
the sequences are more informative than from the fingerprint data.
In other words, 
this approach allows us to order
BACs with higher confidence.  
Also, unlike the use of score functions to attempt to resolve the
repeat problem, our way of resolving the repeat problem 
by considering the consistency of overlaps
and the interval graph formalism is mathematically justifiable.
In addition, this enables us to detect
inconsistencies in the underlying data. We remark that this error
detection feature does not exist in the other algorithms.
Finally, we remark that the idea of our algorithm can be naturally extended
when additional information, such as low-copy repeated sequences are
available. 
A separate work\REF{cshl-meeting} of modifying {\sc Barnacle} to incorporate a  segmental
duplication database\REF{SDD-paper} is in preparation\REF{ROPE-paper}.
\section*{Details of Input}
In this section, we describe the input for {\sc Barnacle}, which
includes the sequence information of BACs and fragments, overlap
information and orientation information.
\subsection*{BACs and fragments}
A {\em BAC} consists of a contiguous stretch of DNA from a chromosome.
The associated data with each BAC consists of the estimated length of
the BAC, the phase of the BAC, the number of fragments, each
fragment's sequence and the chromosome of the BAC, if assigned.

Table~\ref{example-bac} shows an example of  BAC $AC002092.1$ with estimated length
95456bp has four fragments. Fragment $AC002092.1\tilde{\phantom{e}}1$ is
the sequence from 1 to 888 in the GenBank record of $AC002092.1$.
Notice that we do not know the actual sequence of each BAC but we know
the sequence of each of its fragments, which are the
preassemblies of the shotgun reads of the BAC produced by some
assemblers, such as {\sc PHRAP} (Green, unpublished).


Depending on the raw data coverage, BACs are categorized into three phases.
Phases 1 and 2 are the draft
sequences; phase 3 comprises the finished sequences. For a phase 1 BAC, the
fragments of the BAC are not necessarily disjoint, and the order and
orientation of fragments are in general unknown. For a phase 2 BAC, the
fragments are disjoint and the order of fragments is known. A phase
3 BAC is a finished sequence, i.e. only one fragment. In addition,
for some phase 1 BACs, some partial fragment order information and end
fragment information are also available. 

\subsection*{Overlap Information}

Based on local alignments of all fragment sequences against all
fragment sequences, we further process 
two types of
valid overlaps between fragments: dovetail overlap and complete containment
(see Fig.~\ref{valid-overlap-graph}).
The sequence
identity of the overlaps has to be at least $97\%$ and
end-allowed-error (as shown in Fig.~\ref{valid-overlap-graph})
is taken to be $350$bp for
finished sequences and $\min \{\mbox{10\% of the fragment length},
1000\}$bp otherwise.

Also, according to the annotation of some finished sequences, some overlaps
are generated. These are called {\em nt-pairs}. These overlaps include $0$bp
overlaps, i.e., the fragments do not overlap but are consecutive to each
other. The nt-pairs are supposed to be true overlaps if the
annotations are correct and they are generated accordingly.

\subsection*{Orientation Information}
There are additional sequences from paired-end
plasmid reads.
These sequences are
aligned against all fragment sequences. 
Based on these alignments, 
 the relative orientation of fragment pairs is obtained as input.
\section*{Results and the comparison with the public assembly}
In this section, we present the result of our algorithm applied to
the input of the December 2001 freeze of the public working draft of the human
genome.  
First, we describe the details of the
input of the December 2001 freeze.
Then we present 
and compare the results with the NCBI's assembly (Build 28).

\subsection*{Input of the December 2001 freeze}
\label{sec-4.1}
The sequence information with chromosome assignments, local alignments
of all-against-all fragment sequences, local alignments of all
fragment sequences against plasmid reads sequences
were provided by NCBI. We further processed these local alignments to
generate valid fragment pair overlaps and relative orientation of
fragment pairs.  The details of the December 2001 freeze  are shown in
Table \ref{Dec02-input-table}.

\subsection*{Results on the December 2001 freeze}
\label{sec-4.2}
We assemble  $219,031$
non-singleton fragments into $12726$ subcontigs, with length
$2.51$ Gbp.  From these subcontigs, we obtain a clone graph of
$33929$ BACs in $2195$ connected components, which consists of 23145 BACs
in 2052 interval connected components and 1078 BACs in 143 non-interval
connected components.
Upon resolving non-interval components, $139$ problematic BACs, which
are either suspicious chimeras or contains unidentified repeats,  are
removed.
The result is $33722$ non-singleton BACs in $2443$ interval
components, as shown in Table \ref{barnacle-assembly-table}.
We remove $8189$ $(=259230 - 251041 )$ fragments based on our FN
detection which indicates the fragments should be contained in some
subcontigs. 

A graphical display of the assembly of December 2001 freeze of the working
draft of the human genome is available at
(http://www.cs.rutgers.edu/\verb+~+vchoi).

\subsection*{Comparison with NCBI's assembly}
We compare our results with NCBI's assembly on December 2001 freeze (see
Table \ref{ncbi-assembly-table}). The $29537$ $(=251928- 222391)$ fragments were absent from NCBI's build for
reasons we do not know. To further measure the quality of the assembly, we introduce the
definition of {\em warp} $= \frac{\mbox{assembled BAC
    length}}{\mbox{estimated BAC length}}$, which is the ratio of the
assembled BAC length over the estimated BAC length. 
See Table \ref{comparison-table} for the comparison.
There are 1016 BACs in our assembly with assembled BAC length more than 250
Kbp (the longest one is $732,527$ bp), while there are 3786 such BACs (with
the longest one  $19,436,525$ bp) in NCBI's assembly. 
We conjecture that the number
would be even worse if they had not thrown away the 29K fragments.
Note that while
it is true that the warped BACs are
misassembled, it is not necessary that the non-warped BACs are
correctly assembled. Indeed, the suspicious chimeric BACs we detected
are usually non-warped in NCBI's assembly because of the way their
assembly is done. To measure the accuracy of the assembly, it is
important to understand the method or the algorithm of the assembly.
\section*{Discussion}
Our algorithm is very efficient. It takes 3 minutes on a Pentium
III (933 MHz) computer to assemble the  public working draft
of the human genome after preprocessing to the input format as described.
For the same step, {\sc GigAssembler} takes about two hours on a cluster of 100 
Pentium III CPUs. 
We only know it takes less than
one day by NCBI's assembler but do not know the actual time.
Most importantly, unlike the other two algorithms, {\sc Barnacle} is based
on a mathematically justifiable (see section Methods for details) approach for assembling
the sequences. The algorithm not only can detect the inconsistency
of underlying data but also can be well adapted to many situations which
may arise 
as the technology changes and more data becomes available (for example, see
\REF{ROPE-paper}).

Also, we would like to note that another advantage of our algorithmic
approach is 
that we make better use of the available
data.
As it was also pointed out in\REF{Gene2-paper}, the clone-overlaps
inferred from the sequences are more informative than from the
fingerprint data.  While the NCBI algorithm does not make use of the
fingerprint-map as {\sc GigAssembler} does, the way NCBI
makes use of the fragment overlaps to infer the clone overlaps is in
the traditional, less informative physical-mapping clone overlap
fashion.  More precisely, in physical-mapping, one only knows whether
two clones overlap by some common fingerprints, rather than at which
ends they overlap, while such information is explicit in the fragment
overlaps. 
See Figure~\ref{distinction-overlap-graph} for this
important distinction.
In particular, NCBI
formulates a Maximal Interval Subgraph (MIS) problem (which is
NP-hard) to obtain a clone ordering, where the weight of a
clone-overlap is the summation of weights of all the corresponding
fragment pairs, which are assigned by some score functions.  In
contrast, we first ``conservatively'' assemble fragments, which uses
the full information of fragment overlaps (and not just whether two
fragments overlap), and then infer the clone overlaps from these
subassemblies. 

It is worthwhile to point out that it is theoretically impossible to 
assemble the true assembly based only on the sequence overlap information,
although
the way we resolve the repeat problem is justified by considering the consistency
of overlaps and the interval graph formalism.
Indeed the assemblies are far from perfect.  For example, the
assemblies of those ``warped'' BACs whose assembled length is greater
than 250Kb are obviously incorrect. These misassemblies might be
largely due to undetected FNs, or they might be due to over-collapse
of repetitive regions, which do not destroy the interval graph
property; 
for example,
repeats which occur at one end of contigs due to the lack of coverage
or repeats which are longer than an entire BAC.  More biological
information is needed to resolve these cases. For instance, the 500Kb
inverted repeat on chromosome Y is
resolved by using the annotation of sequences in the GenBank.
Also, 
the detected suspicious errors, which include chimeras, chromosome
misassignments, wrong annotation of some finished sequences and
fragment misassemblies need further follow-up and verification by the
sequencing centers.  We believe that in order to achieve a
high-quality assembly, an iterative process involving 
collaboration with the sequencing centers is necessary.

An original goal of the Human Genome Project is to provide an accurate
reference sequence of the euchromatic portions of all human
chromosomes\REF{Collins-paper} and it is essential for an assembler
to resolve repeats correctly in order to achieve accurate
results\REF{care-paper}.  
It is our hope that future (automated) assemblers
for clone-based sequences of whole genomes (including other higher
organisms, such as mouse, maize and rice etc.) will be developed and
improve on this well justified framework. 
Since the human genome project, 
new sequencing strategies have
been developed, e.g., hybrid of clone-based and whole-genome shotgun.
Nevertheless, we believe researchers can benefit from the algorithmic
skills to develop the new assemblers.
The source code of {\sc Barnacle} is freely
availabe at (http://www.cs.rutgers.edu/\verb+~+vchoi). 
\section*{Methods}
Before we describe the idea behind our algorithm, we introduce some
terminology.

We say an overlap is {\em true} if the fragments are from the
overlapping segments of the genome, otherwise the overlap is {\em
repeat-induced} 
(Fig.~\ref{overlap-graph})\REF{Gene-paper}.
The overlaps which are not true are called {\em False Positives}
(FPs). In other words, repeat-induced overlaps are FPs. On the other
hand, the overlaps which are true but not detected are called {\em
False Negatives} (FNs). FPs and FNs are called {\em noise}. 

The assembly problem  would be straightforward if we could divine all true
overlaps, i.e., if the data of overlaps were noise-free. The key objective
is thus to clean up the noise as much as possible and assemble the
fragments according to the true overlaps.

Observe that for an
overlap to be true, it is necessary for the overlap to be
``consistent''
(Fig.~\ref{consist-vs-inconsist-graph}).
However, ``consistency'' is not
sufficient for a true overlap. There might be some consistent  
but repeat-induced overlaps, which
are due to either the lack of coverage or long repeats
(Fig.~\ref{nonconflict-overlap-graph}).
We call a consistent but repeat-induced overlap a {\em consistent
repeat-induced overlap}, otherwise we call it an {\em inconsistent repeat-induced overlap}.

A {\em contig} (subcontig resp.) is a contiguous region that is covered by a set of
overlapping BACs (fragments resp.). 

\label{sec-3.2}
The basic idea behind our algorithm is illustrated in
Figure~\ref{simple-idea-graph} and Table~\ref{high-level-alg} shows the
high level description of the algorithm. In 
following, we outline the idea of each step.
The details of the implementation of the algorithm are described in
\REF{PhD-thesis}(http://www.cs.rutgers.edu/\verb+~+vchoi).

\begin{description}
\item {\bf ``Conservatively'' assemble fragments into
subcontigs.}
First, we assemble consistent overlapping fragments into
subcontigs. Before we describe the algorithmic implementation, we
 introduce some terminology. A fragment is  called a {\em subfragment}
if the fragment is completely contained in another fragment, otherwise the
fragment is {\em maximal}. Making use of the maximality property of the
latter, we efficiently identify and assemble consistent overlapping maximal
fragments (once again readers are invited to read \REF{PhD-thesis} for the
simple algorithmic trick). Then consistent overlapping subfragments are 
assembled into the contig of their corresponding maximal fragments. 
With this implementation, 
$219,031$ non-singleton fragments of the December 2001 freeze are assembled 
into $12,726$ subcontigs in about
25 seconds on a Pentium III computer.

We then deduce clone-overlaps from these subcontigs: two
clones overlap if and only if there is at least one fragment pair of
the corresponding clones overlapping in a subcontig.
Then the conflicting overlaps are resolved according to the
clone-overlaps (Fig.~\ref{resolve-inconsist-graph}). 
Unlike the use of score functions to resolve conflicts, we are making use of
the BAC information of fragments and use the assembly obtained by
consistent overlaps to resolve inconsistent overlaps. This approach is well
justified because the consistent overlaps, which are necessary condition
for true overlaps, give a good indication whether two BACs overlaps or not.  
Note that it is this approach that allow us to naturally make use of the segmental
duplication database without substantial changing the algorithm.

As mentioned before, these subcontigs might still contain some
consistent repeat-induced overlaps.
\item{\bf Detect and remove consistent repeat-induced overlaps and chimeric
clones.}
We use the linear structure of the chromosome to 
detect the consistent repeat-induced overlaps.
The linear structure of the sequence would be destroyed if the
repeat-induced overlaps were used (Fig.~\ref{noninterval-graph}{\bf a}).
Mathematically speaking, the corresponding clone graph, whose vertices
are clones and there is an edge between two vertices if and only if
the two corresponding clones overlap, must be an interval
graph. (Remark: This is true with the assumption that the length of gaps
  within a BAC is short enough such that there
  is at least one fragment pair overlapping if two BAC overlaps. However,
  this assumption is no longer true in the current working draft of the
  human genome because some BACs were heavily trimmed resulting some of the
   BACs only a few hundreds bp long (much shorter than a
  possible gap). Comparing
   April 2001 freeze with December 2001 freeze,
  we found that 1200 BACs whose length have been reduced more than half of
  its original length. Note that our algorithm is robust enough that
  with a slight modification of resolving non-interval graph procedure, one can 
  heuristically take care of this possibility.
)
Intuitively, imagine
the genome sequence as a line, then each BAC is an interval of the
line such that two BAC sequences overlap if and only if the
corresponding intervals overlap (Fig.~\ref{interval-graph}{\bf a}).
In other words, if the corresponding clone graph is not interval,
 the assembly is incorrect.
For example, BAC AC019248 in NCBI's Build 28 is misassembled (Fig.~\ref{interval-graph}{\bf b}).
By resolving the non-interval connected components of the clone 
graph,
we detect and remove suspicious repeat-induced overlaps and chimeric clones
(Fig.~\ref{noninterval-graph}{\bf b}).
Instead of resorting to the NP-hard Maximal Interval
Subgraph problem, which does not characterize the real biology, to get an
interval graph, we design an efficient algorithm for resolving the
non-interval graph (see Table~\ref{rn-table} for the idea).
The interval representation (and hence the ordering) of clones is
obtained from the resulting interval clone graph by a
linear-time interval graph recognition algorithm \REF{5-sweep-paper}. 

\item {\bf Orient and order subcontigs.} 
According to the interval representation of
clones, 
subcontigs are
ordered and 
the orientation of ``long'' subcontigs is determined. 
According to the ordering of clones,  we orient subcontigs by flipping such
that the ranks of BACs in each subcontig are in non-decreasing order, and
assign coordinates to subcontigs so that they can be ordered by sorting
according to these coordinates lexicographically. 
Note that our interval representation of clones is quite informative 
in that most of the subcontigs can be ordered unambiguously according to this
 reliable information. Thus we do not need to employ the quite
 noisy and uncertain information from plasmid reads, ESTs and mRNAs to
 order the
 subcontigs, as does {\sc GigAssembler}, which resorted to the
 Bellman-Ford algorithm for testing the feasibilities of the
 information.
Also, the interval representation allows us to detect FNs while
ordering subcontigs.
First observe that the corresponding end BACs of
the adjacent subcontigs must be either the same or overlapping
(Fig.~\ref{adjacent-condition-graph}{\bf a}). We call this necessary
condition the {\em adjacency condition}.
Accordingly, when some subcontigs cannot be ordered such that they satisfy the
adjacency condition, it indicates that there might be FNs (Fig.~\ref{adjacent-condition-graph}{\bf b}). 
To
further verify the identification of the FNs, we aligned the involved fragments with their
overlapping clones. Examining these alignments reveals the several possible
causes, which include the consequences of 
repeat-masking, low accuracy of some draft sequences, chimeric
fragments or fragment misassemblies  and  polymorphism.
On the other hand, some subcontigs might be {\em ties} in which the adjacency
condition is always satisfied for any permutation of them. In other words, the
information is not sufficient to determine their order. In this case, we employ
some additional information to break these ties.

\item{\bf Adjust the ordering and correct the orientation of the subcontigs
using additional information.} The additional information
includes the
identification of the end fragments of
BACs, as well as, the partial order of some fragments. 
For each BAC which has end fragment information, i.e., one or two
fragments, first according to their current position in the contig, we
determine which one is the left end fragment and which one is the right end
fragment. Then we orient the fragments so that they are the extreme
fragments of the BAC. To ensure the reliability of the information, these
adjustments (order and orientation) are always subject to the adjacency
condition, i.e., whether we can adjust the order and orientation according
to the information such that the adjacency condition is still satisfied.
Similarly, the adjustments are being done for each group of the ordered fragments.

Finally, the relative
orientation of fragment pairs generated from plasmid reads are
used to orient the subcontigs which are not ``long'' enough to have
been determined. 
According to the interval representation of clones, one can determine how
confident the orientation of subcontigs is. For example, for the subcontigs
which are long enough so that the smallest BAC and the largest BAC are not
overlapping in the subcontig, the correctness of the orientation is sure. The relative
orientation of fragment pairs generated from plasmid reads are
used to orient the unsure subcontigs. For each unsure subcontig, we
organize all its relative orientations into a list of agreeing subcontigs
and a list of disagreeing subcontigs. We then progressively change the
orientation of unsure subcontigs such that the number of disagreeing
subcontigs is minimized. 

\end{description}

\section*{Acknowledgements}
Vicky Choi was supported by a Fellowship from the Program in
Mathematics and Molecular Biology (PMMB) 
at the Florida State University, with funding from the Burroughs Wellcome Fund Interfaces Program
and 
partially supported by NIH Supplemental Visiting Fellowship while
visiting NCBI. We acknowledge NCBI for introducing the problem and providing
access to the data. 
We would like to thank Navin Goyal, JinSheng Lai, Knut Reinert,
Granger Sutton, Wojciech Makalowski, Dannie Durand and  Sachin Lodha   
for their critical reading and comments. Special thanks to Craig
Nevill-Manning for providing the Pentium III computer to finish up the
work, Joe Nadeau for the encouragement
and Evan Eichler for the comments and advice.

\newpage
\section*{Figure Legends}

\begin{figure}[h]
\doublespacing
\begin{center}
\includegraphics[width=2.5in]{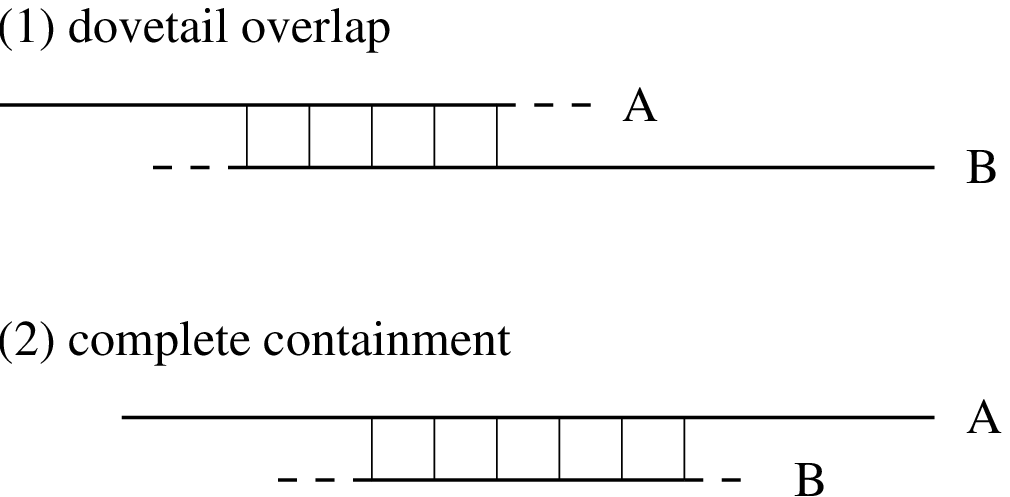}
\caption{}
\label{valid-overlap-graph}
\end{center}
\end{figure}

\noindent
Figure~\ref{valid-overlap-graph}:
Two types of valid overlaps between a fragment pair. (1) dovetail overlap vs. (2) complete
containment. Because of the draft quality of sequences, end errors are
allowed (shown as dashed lines).

\begin{figure}[h]
\doublespacing
\begin{center}
\includegraphics[height=3in]{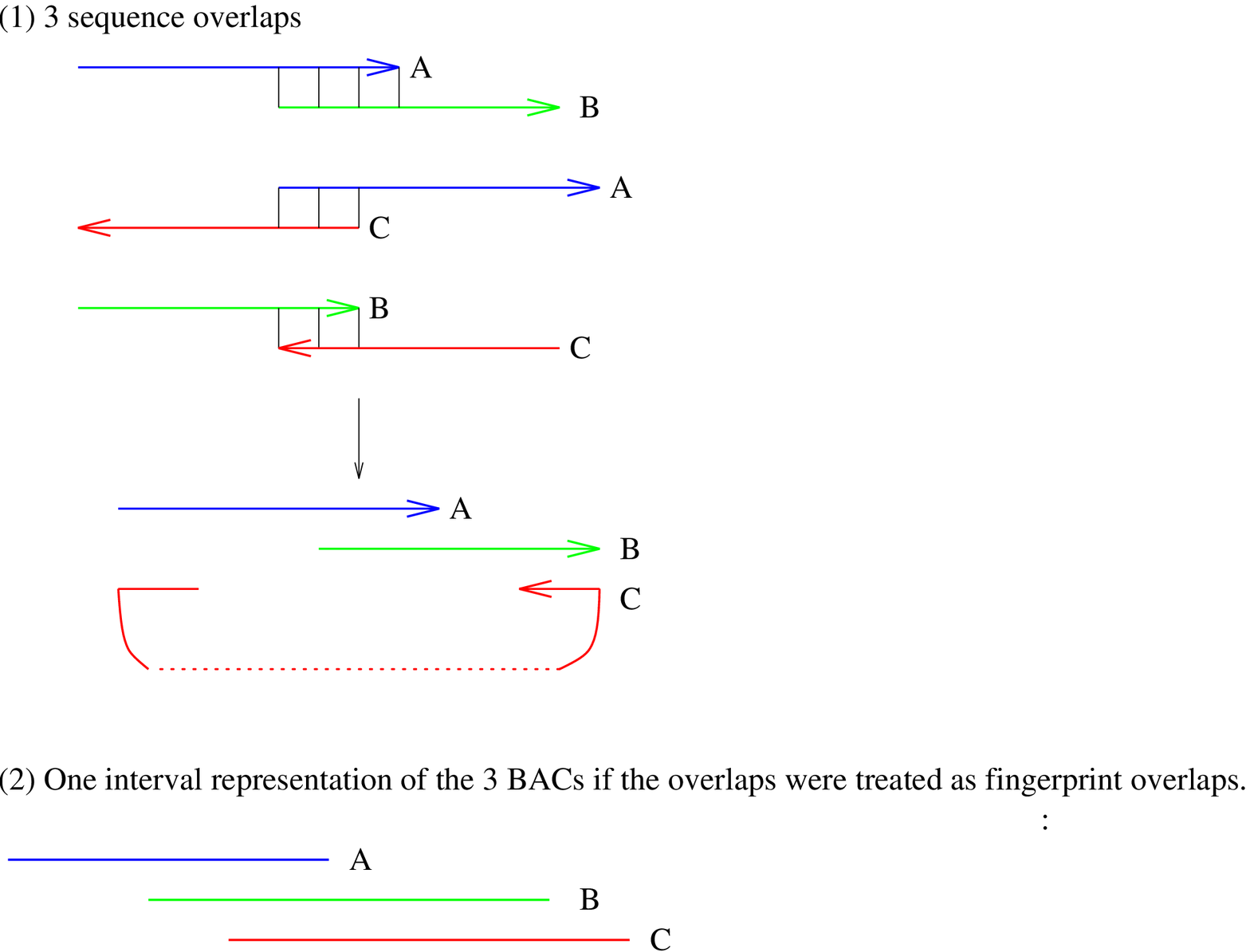}
\caption{}
\label{distinction-overlap-graph}
\end{center}
\end{figure}

\noindent
Figure~\ref{distinction-overlap-graph}:
Sequence overlap vs. fingerprint overlap.
(1) There are 3 sequence overlaps, but they are incompatible. At least one of
them is FP. (2) Treated as fingerprint overlaps (whether two fragments
overlap) would result in an incorrect interval representation of
the 3 BACs.

\begin{figure}[h]
\doublespacing
\begin{center}
\includegraphics[width=4in]{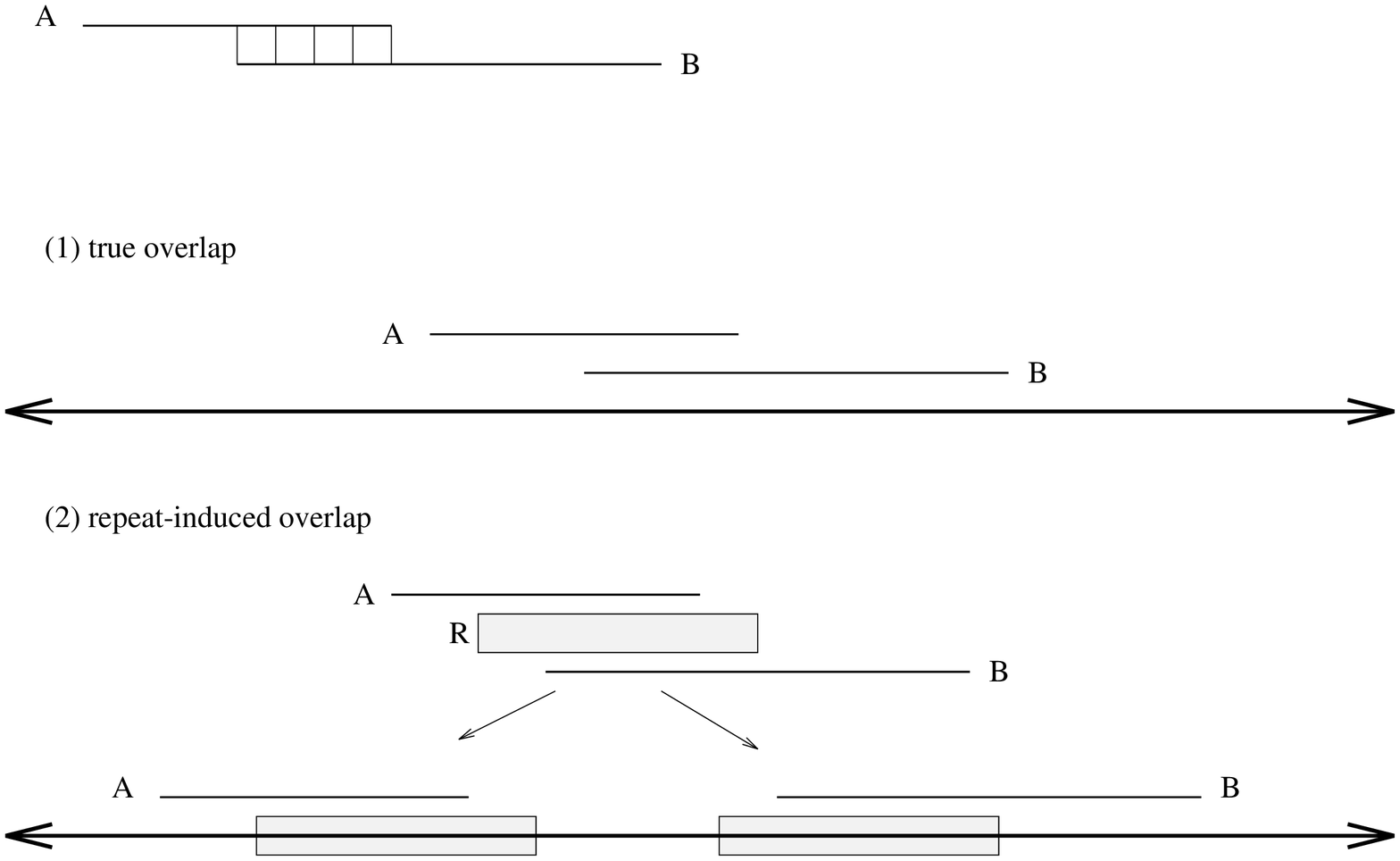}
\caption{}
\label{overlap-graph}
\end{center}
\end{figure}

\noindent
Figure~\ref{overlap-graph}:
True vs. repeat-induced overlaps. Consider the overlap of two
fragments A and B. There are two possible inferences: (1) The overlap is true,
where the fragments are from the overlapping segments of the
genome. (2) The overlap is induced by repeated sequence R.

\begin{figure}[h]
\doublespacing
\begin{center}
\includegraphics[width=4in]{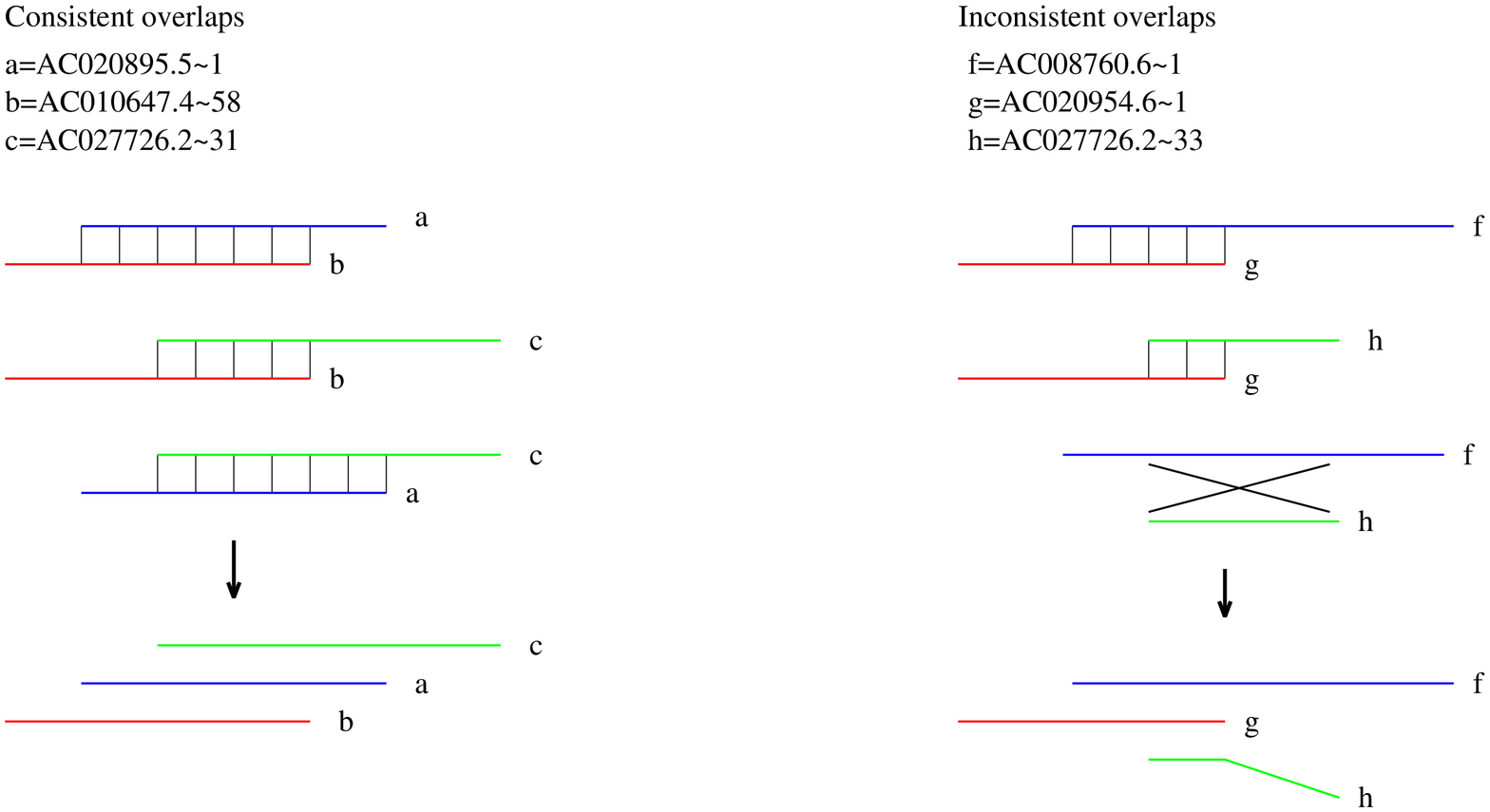}
\caption{}
\label{consist-vs-inconsist-graph}
\end{center}
\end{figure}

\noindent
Figure~\ref{consist-vs-inconsist-graph}:
Consistent vs. inconsistent overlaps.  
Left, fragment $b$ overlaps with fragments
$a$ and $c$ as shown, and
$a$ and $c$ also overlap accordingly. 
The overlap of $b$ and $a$ is {\em
consistent}.
Right, fragment $g$ overlaps with fragments 
$f$ and $h$ as shown, but 
$f$ and $h$ do not overlap accordingly.
The overlaps of $g$ and $f$, $g$
and $h$, are {\em inconsistent}. At least one of them is
repeat-induced.

\begin{figure}[h]
\doublespacing
\begin{center}
\includegraphics[width=4in]{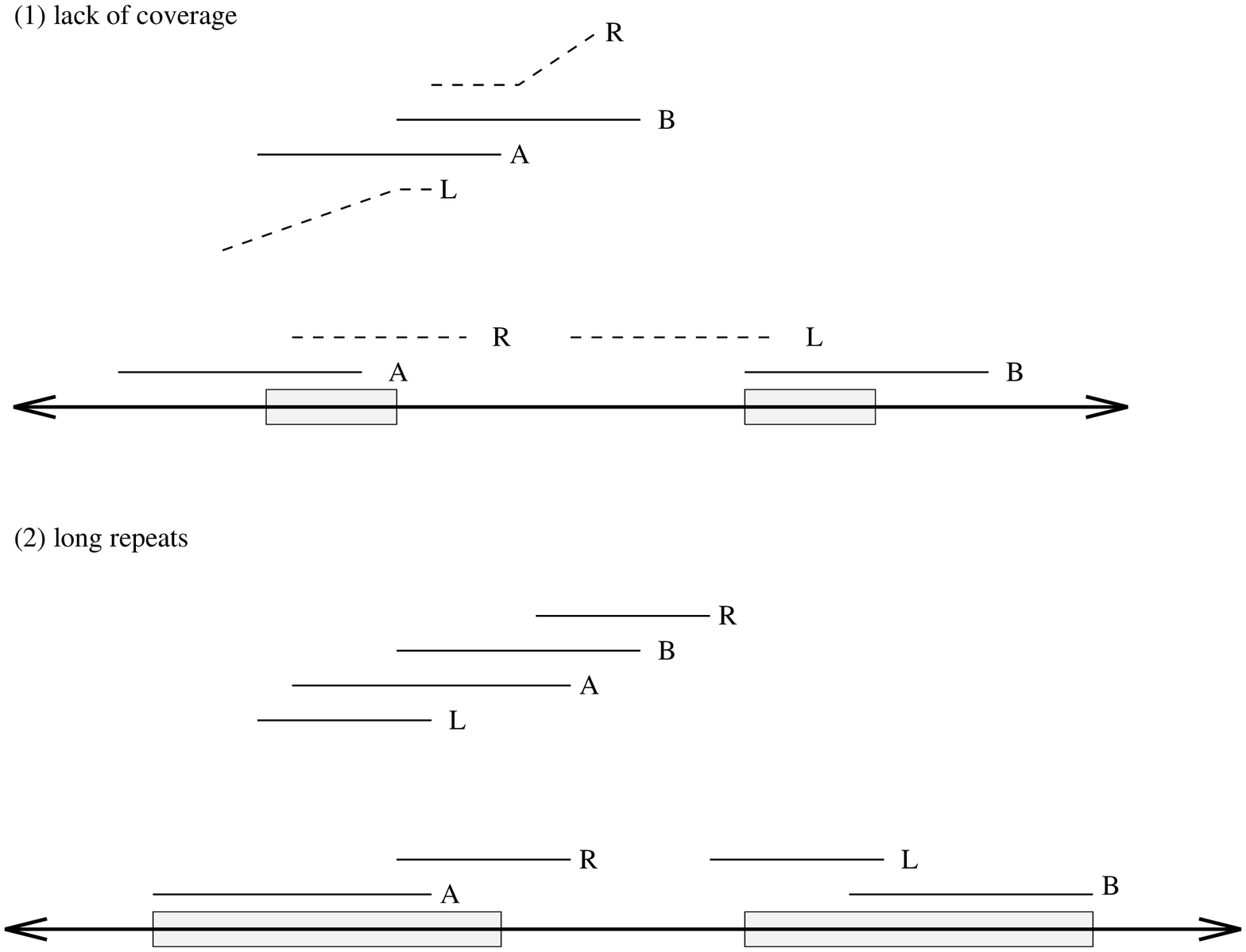}
\caption{}
\label{nonconflict-overlap-graph}
\end{center}
\end{figure}

\noindent
Figure~\ref{nonconflict-overlap-graph}:
Overlaps that are consistent but not true. There are two
possibilities: (1) lack of coverage: absence of both fragments L and
R; (2) long repeats: the repeat is long enough such that the overlaps
are consistent.

\begin{figure}[h]
\doublespacing
\begin{center}
\includegraphics[width=4in]{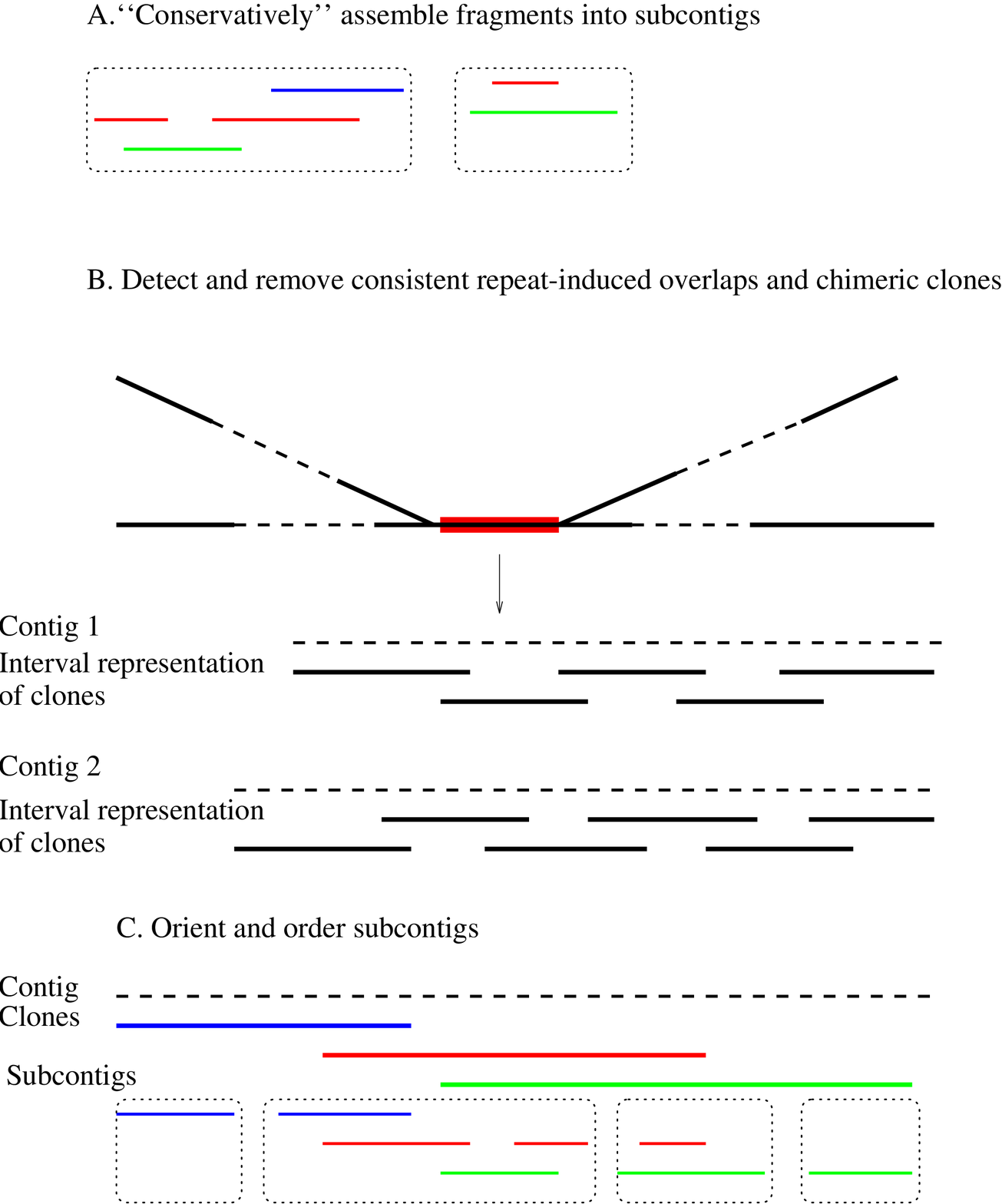}
\caption{}
\label{simple-idea-graph}
\end{center}
\end{figure}

\noindent
Figure~\ref{simple-idea-graph}:
The idea behind our algorithm. Conceptually, the algorithm
consists of three steps. 
(A) Fragments are ``conservatively'' assembled into subcontigs. The
details of this step constitute steps 1-6 of the algorithm.
(B) Consistent repeat-induced overlaps and chimeric clones will
destroy the interval property of the clone graph. Resolving this step
corresponds to step 7 of the algorithm. 
(C) According to the interval realization of clones 
 obtained from the interval
clone graph, subcontigs are oriented and ordered in steps 8-10 of 
the algorithm. Remark: Steps 11-15 of the
algorithm, which use additional information to adjust the ordering and
correct the orientation, are not shown in this figure.

\begin{figure}[h]
\doublespacing
\begin{center}
\includegraphics[width=3in]{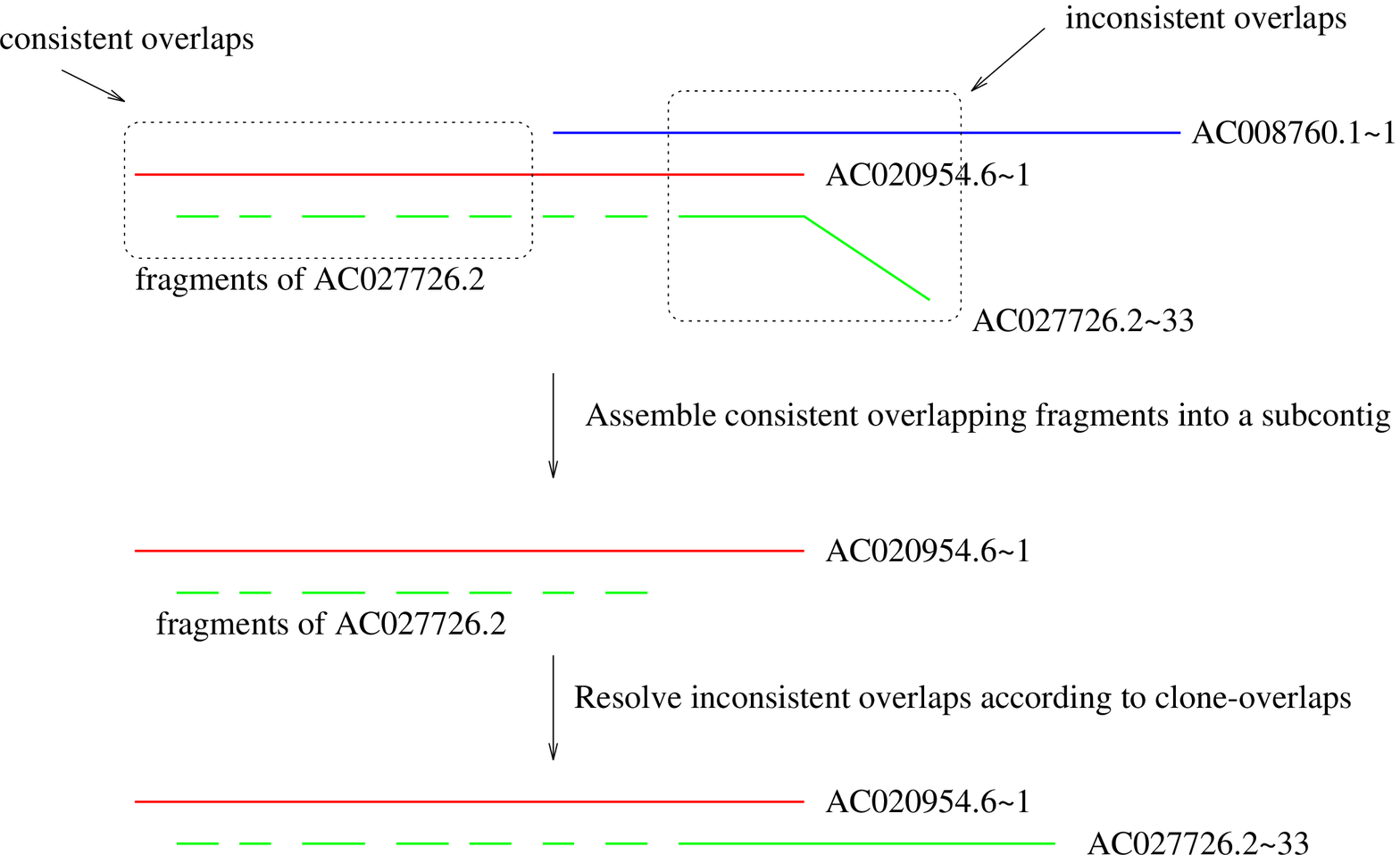}
\caption{}
\label{resolve-inconsist-graph}
\end{center}
\end{figure}

\noindent
Figure~\ref{resolve-inconsist-graph}:
Resolving inconsistent overlaps. 
Since 
$AC020954.6$\~{}1 has several consistent overlaps with fragments of BAC
$AC027726.2$, BAC $AC020954.6$ overlaps with BAC $AC027726.2$.
The inconsistent overlaps of $AC020954.6$\~{}1 and $AC008760.6$\~{}1, $AC020954.6$\~{}1
and $AC027726.2$\~{}33 are resolved according to the clone-overlap. In this case,
the overlap of $AC020954.6$\~{}1
and $AC027726.2$\~{}33 is chosen, i.e. the overlap  of $AC020954.6$\~{}1 and $AC008760.6$\~{}1
is repeat-induced.

\begin{figure}[h]
\doublespacing
{\bf a}\\
\includegraphics[width=4in]{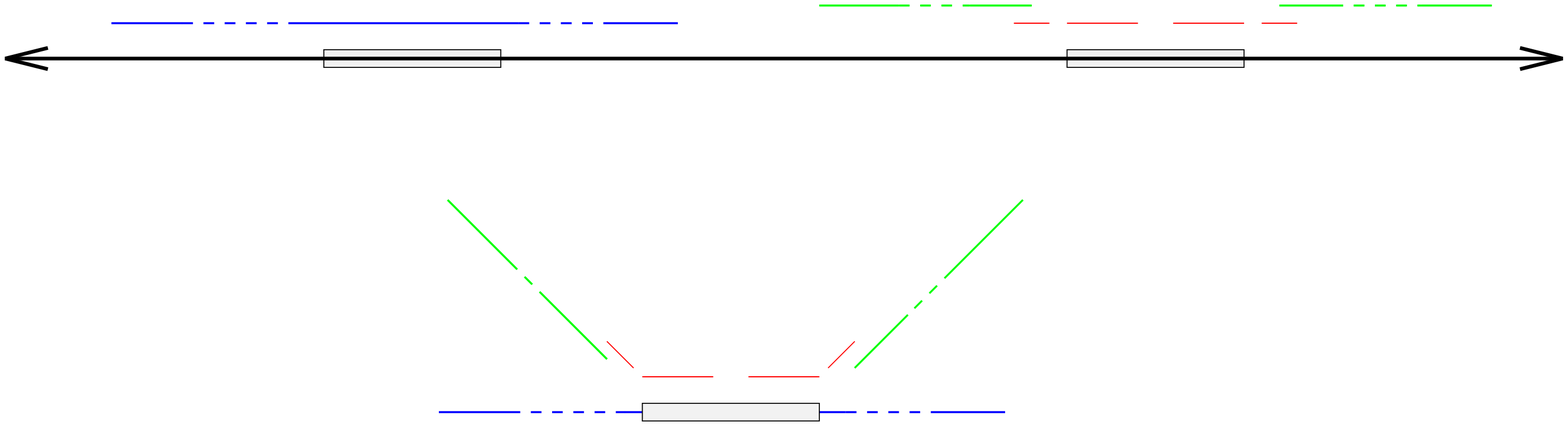}

\vspace*{0.3cm}
{\bf b}\\
\includegraphics[width=4.5in]{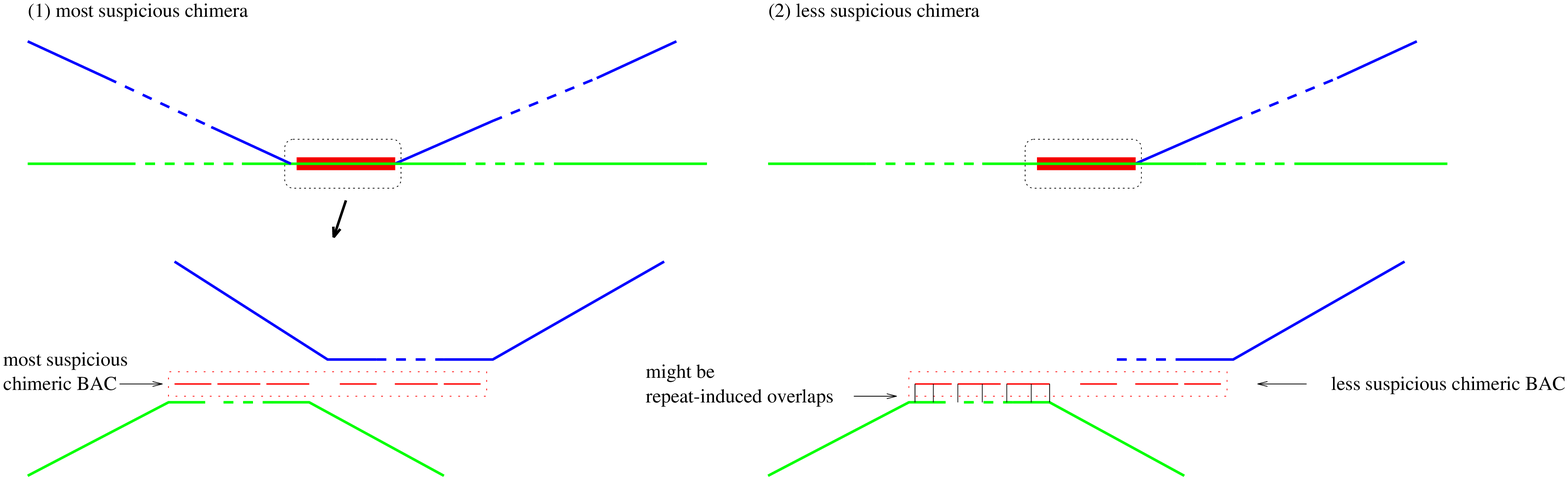}
\caption{}
\label{noninterval-graph}

\end{figure}

\noindent
Figure~\ref{noninterval-graph}:
{\bf a}
The result of collapsing a repeat region. The blue and green  
denote contigs covered by a set of overlapping BACs; while the red
represents a BAC consisting of 4 fragments.  For illustrative
purposes, suppose the repeat region occurs in the two fragments of the
red BAC. The linear structure of the sequence would be destroyed if
the repeat region were to be collapsed.  
{\bf b}
Suspicious chimeric BACs. Left, 
the problematic region occurs in the
middle of a pair of contigs, the BAC is the most suspicious chimera.
Right, when the problematic region occurs at
one end of one contig, it is difficult to tell that it is due to a
repeat or whether
the BAC is chimeric. To ensure the quality, the BAC is removed.

\begin{figure}[h]
\doublespacing

{\bf a}\\
\includegraphics[width=4in]{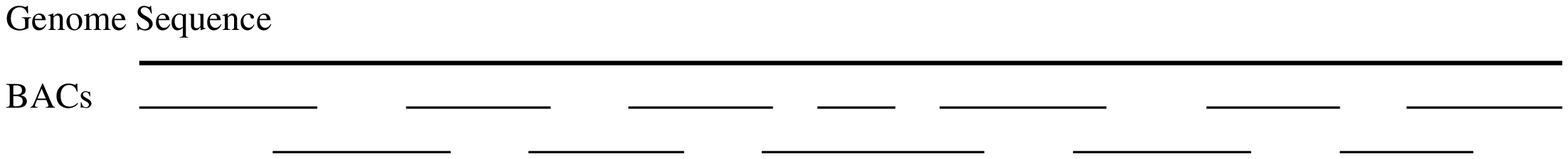}

\vspace*{0.3cm}
{\bf b}\\
\includegraphics[width=5in]{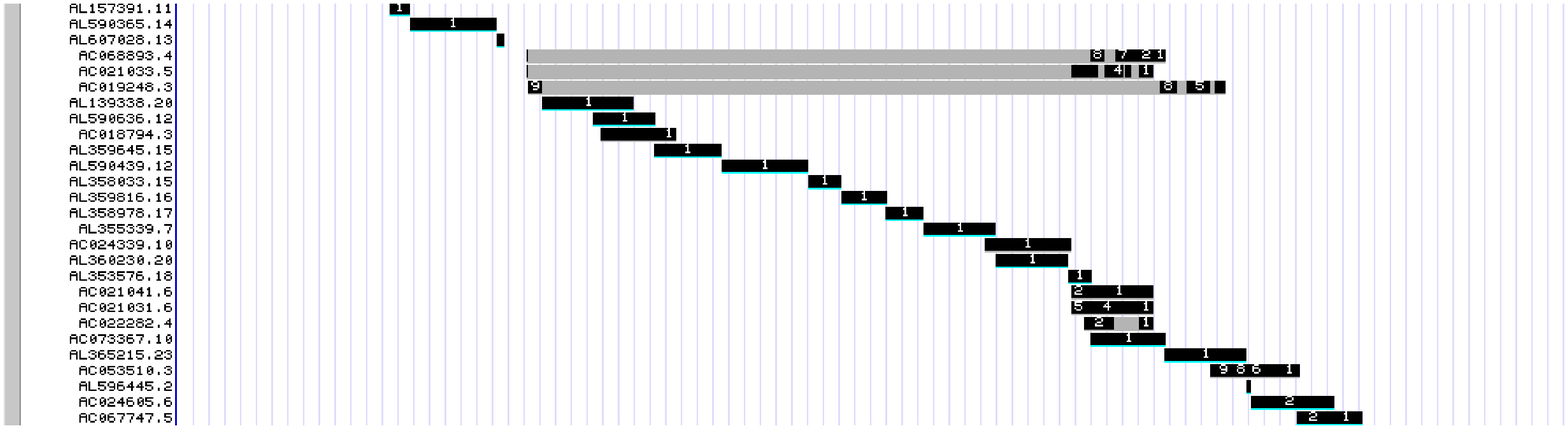}
\includegraphics[angle=270,width=5in]{AC019248.epsi}
\caption{}
\label{interval-graph}

\end{figure}

\noindent
Figure~\ref{interval-graph}:
{\bf a} Geometric view of the sequences. The genome sequence is a line; each BAC
corresponds to an interval of the line.
{\bf b} 
Above, the contig of BAC $AC019248$ in NCBI's Build 28, which consists of 10 fragments (only 9 fragments were
used in NCBI's Build 28), is non-interval.
Biologically, the misassembly of BAC $AC019248$ is also evident by its
assembled length of $1,556,292$bp.
Below, {\sc Barnacle}'s interval-graph assembly of the corresponding contig.

\begin{figure}[h]
\doublespacing
{\bf a}\\
\includegraphics[width=4in]{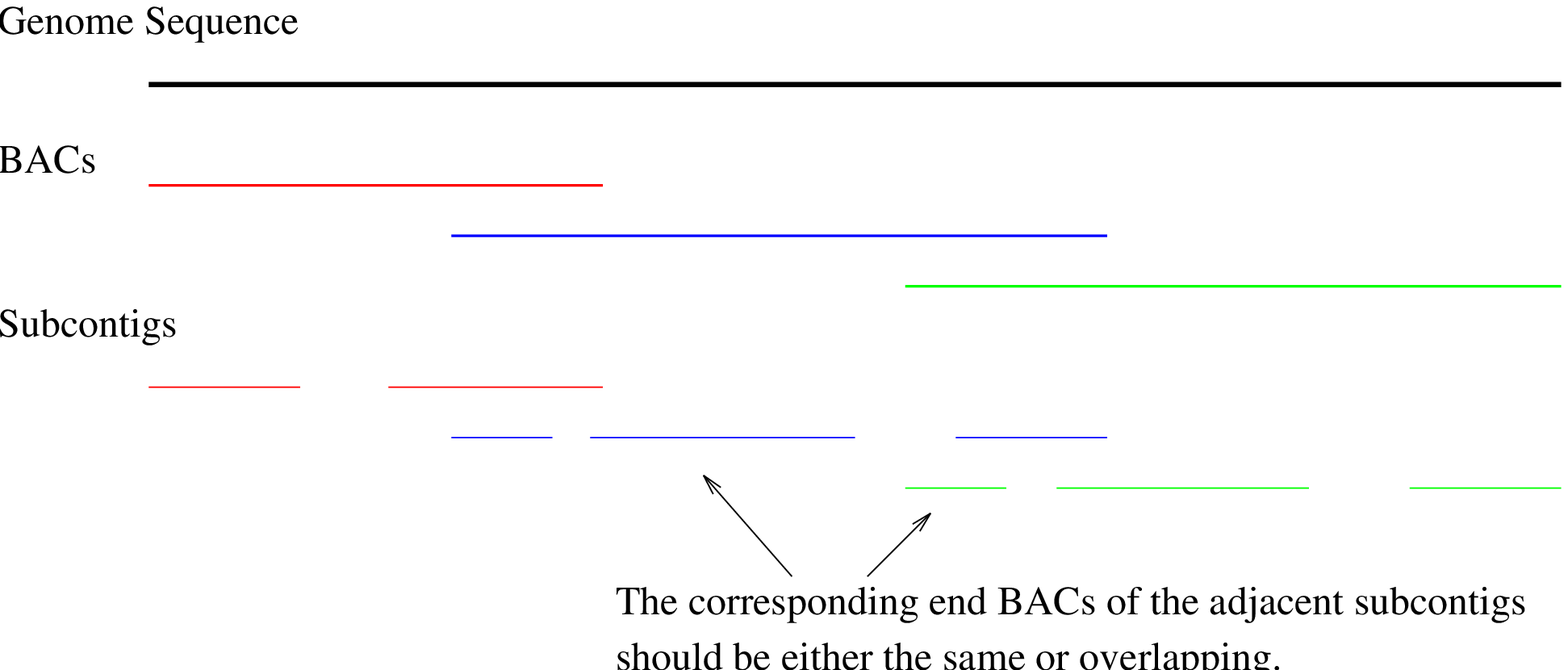}

\vspace*{0.3cm}
{\bf b}\\
\includegraphics[width=4in]{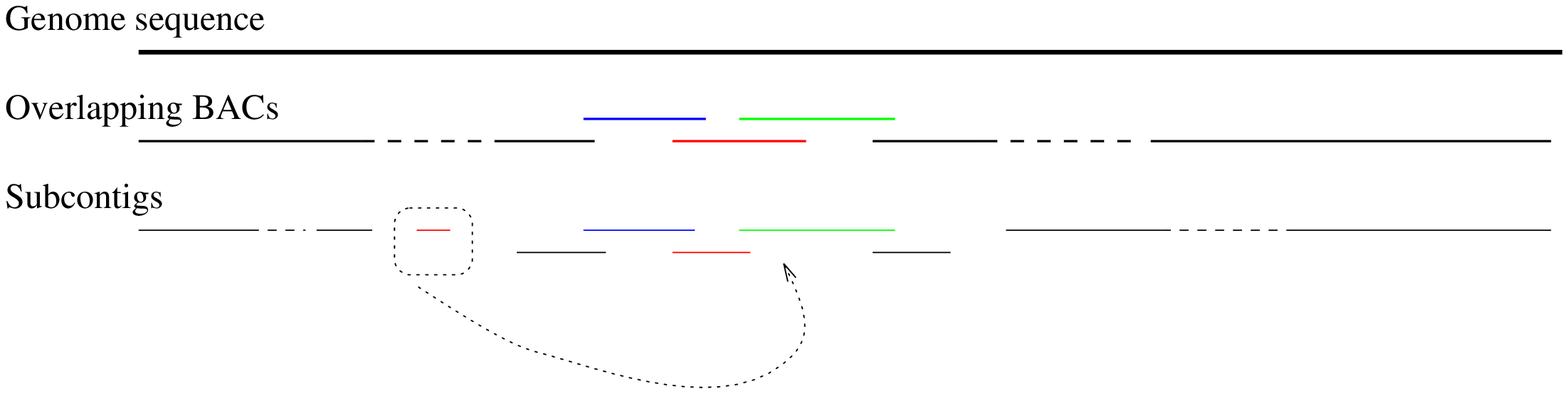}
\caption{}
\label{adjacent-condition-graph}

\end{figure}

\noindent
Figure~\ref{adjacent-condition-graph}: {\bf a}
The condition of adjacent subcontigs. The corresponding end BACs of
the adjacent subcontigs must be either the same or overlapping.
{\bf b} 
FNs detection. No matter how we order the subcontigs, the
subcontig in the box will violate the adjacency condition. This is
due to a FN as the arrow shows.


\newpage

\newpage


\newpage

\newpage

\newpage

\newpage

\newpage

\newpage

\newpage

\newpage

\renewcommand{\arraystretch}{1.5}
\newpage
\begin{table}[h]
\begin{verbatim}
accession #   estimated length  phase  chrm  number of fragments
AC002092.1      95456             1     17      4
        fragment acc#   start   end     length
        AC002092.1~1    1       888     888
        AC002092.1~2    889     46200   45312
        AC002092.1~3    46201   84925   38725
        AC002092.1~4.1  84926   95170   10245
\end{verbatim}
\caption {An example of a BAC.}
\label{example-bac}
\end{table}

\begin{table}[b]


\begin{enumerate}
\item{\bf Sequence Information}
$$
\begin{array}{crrcc}
\mbox{phase}&\mbox{BACs}&\mbox{fragments}&\mbox{total length in Gbp}&\mbox{average number of fragments}\\
\hline
1&	15298&	246424&		2.55&		16.11\\
2&	 2154&	  8161&		0.33&		 3.79\\
3&	17624&	 17624&		2.04&		 1.00\\
\hline
\mbox{Total}&	35076&	272209&		4.992&		 7.76
\end{array}
$$

\item {\bf Chromosome Assignment}

The scheme employed by NCBI to assign a chromosome to a BAC is based
on:
\begin{itemize}
\item STS: presence of at least 2 STS markers that have themselves 
been mapped to the same chromosome; 
\item GenBank: annotation on the submitted GenBank record;
\end{itemize}
Otherwise, the chromosome of the BAC is {\em Unknown}.

The chromosome assignments are summarized as below: $$
\begin{array}{ccc}
\mbox{STS}&\mbox{GenBank}&\mbox{Unknown}\\
\hline
31543 &2450 &1083
\end{array}
$$

\item{\bf Overlap Information}
There are $403,466$
fragment pairs of the same chromosome or at least one of them an
unknown chromosome. There are $12,656$ nt-pairs.

\item{\bf Orientation Information}
Relative orientation fragment pairs are generated from paired-end
plasmid reads. There are $321,751$ such fragment pairs.

\end{enumerate}
\caption{The details of the input  of the December 2001 freeze.}
\label{Dec02-input-table}
\end{table}

\begin{table}[h]
$$
\begin{array}{lcccc}
&\mbox{BACs}&\mbox{Fragments Used/Fragments}&\mbox{Contigs}&\mbox{Length in }Gbp\\
\hline
\mbox{singletons}&1215&9967/9967&1215&0.142\\
\mbox{non-singletons}&33722&251041/259230&2443&2.708\\
\hline
&34937&261008/269197&3658&2.850
\end{array}
$$
\caption{The results of {\sc Barnacle} on December 2001 freeze.}
\label{barnacle-assembly-table}
\end{table}

\begin{table}[h]
$$
\begin{array}{lcccc}
&\mbox{BACs}&\mbox{Fragments Used/Fragments}&\mbox{Contigs}&\mbox{Length in }Gbp\\
\hline
\mbox{singletons}&836&9074/9074&836&0.112\\
\mbox{non-singletons}&32902&222391/251928&2292&2.745\\
\hline
&33738&231465/261002&3128&2.857
\end{array}
$$
\caption{Build 28 - NCBI's assembly on December 2001 freeze.}
\label{ncbi-assembly-table}
\end{table}

\begin{table}[h]
$$
\begin{array}{ccc}
\mbox{Warp} & \mbox{Our Assembly} & \mbox{NCBI's Public Assembly}\\
\hline
<=1.5 &33474& 29647 \\
1.5 - 1.8 & 753 & 725\\
1.8 - 2.0 & 278& 371\\
2.0 - 5.0 & 421& 1813\\
5.0 - 10.0& 10 & 612\\
>10.0 & 1 & 570
\end{array}
$$
Restricting 
warp $>1.5$, 
$$ 
\begin{array}{lcc}
\mbox{Assembled BAC Length} & \mbox{Our Assembly} & \mbox{NCBI's
Public Assembly} \\
\hline
250K - 300K& 434&461\\
300K - 500K& 549&1328\\
500K - 800K& 33&798\\
800K - 1M & 0&248\\
1M - 2M&0&496\\
2M - 3M&0 & 129\\
3M - 10M&0&259\\
10M - 20M&0 & 67 \\
\hline
\mbox{Total} & 1016 &3786
\end{array}
$$
\caption{Comparison of {\sc Barnacle}'s assembly with NCBI's assembly on
December 2001 freeze.}
\label{comparison-table}
\end{table}

\begin{table}[h]

\begin{tabular}{l}
A. {\bf ``Conservatively'' assemble fragments into
subcontigs.}\\
\quad 1.~Classify fragments: singleton, subfragment and maximal
fragment.\\
\quad2.~Assemble consistent overlapping maximal fragments into subcontigs.\\
\quad3.~Put back {\em good} subfragments to subcontigs.\\
\quad4.~Detect and resolve conflicting chromosome assignments.\\
\quad5.~Construct a BAC graph from subcontigs.\\
\quad6.~Resolve inconsistent overlaps according to the connectivity of the
BAC graph.\\
B. {\bf Detect and remove consistent repeat-induced overlaps and chimeric
clones.}\\
\quad7.~For each component $G_i$ of the BAC graph, if $G_i$ is not
interval, resolve the \\\qquad component
 by removing repeat-induced overlaps or
suspicious chimeric BACs. \\
C. {\bf Orient and order subcontigs.}\\
\quad8.~Obtain the interval representation of BACs. \\
\quad9.~Orient subcontigs. \\
\quad10.~Assign coordinates to subcontigs and order subcontigs by
sorting lexicographically. \\
\quad11.~Detect the potential false negatives and remove the involved
fragments. \\
\quad12.~Detect false positives (consistent repeat-induced overlaps that
do not destroy \\\qquad the interval property).\\
D. {\bf Adjust the ordering and correct the orientation of the subcontigs
using} \\{\bf additional information.}\\
\quad13.~Adjust the order of the subcontigs according to the extra fragment
information.\\
\quad14.~Orient subcontigs according to the relative orientation 
of fragment pairs  generated \\\qquad from 
plasmid reads, ESTs and mRNAs data.\\
\quad15.~Derive a consensus sequence for each contig from the assembly
of maximal  \\\qquad fragments of the contig.
\end{tabular}

\caption{The high level description of the algorithm.}
\label{high-level-alg}
\end{table}

\begin{table}[h]

In the following, we sketch the idea of how the non-interval
graphs are resolved.
The interested reader is invited to
read~\REF{PhD-thesis} for the details.

Let $G=(V,E)$ be a non-interval graph. Without lose of generality,
assume that $G$ is connected.
%
%
Define a vertex $v\in V$ to be {\em I-critical} if
$G|_{V\setminus\{v\}}$ is interval.

Given a non-interval graph $G$, we first identify a forbidden subgraph
of $G$, then check whether at least one of the vertices of the
forbidden subgraph is I-critical, if so, we say $G$ is {\em fixable}.

Based on the structure of the forbidden subgraph, a fixable graph $G$
is resolved by 
\begin{enumerate}
\item adding an FN edge; or
\item removing FP edges due to an identified repeat; or
\item removing a vertex which is either a suspicious chimera or
contains an unidentified repeat. 
\end{enumerate}

For the non-fixable graphs, we employ a divide-and-conquer method by
dividing the graph according to some articulation points such that
each subcomponent is fixable. After fixing the subcomponents, the
non-fixable graph would become fixable because the articulation point
would become I-critical.
\caption {The idea of resolving non-interval graph.}
\label{rn-table}
\end{table} 
\end{document}